\documentclass[
superscriptaddress,
twocolumn,
showpacs,preprintnumbers,
amsmath,amssymb,
aps,
prb,
]{revtex4-2}
\usepackage[colorlinks,
linkcolor=blue, urlcolor=blue, anchorcolor=blue, citecolor=blue]{hyperref}
\usepackage{graphicx}
\usepackage{dcolumn}
\usepackage{bm}
\usepackage{float}
\usepackage{multirow}

\newcommand{\angstrom}{\textup{\AA}}
\begin{document}


\title{Magnetic frustration in the cubic double perovskite Ba$_2$NiIrO$_6$}


\author{Ke Yang}
\affiliation{College of Science, University of Shanghai for Science and Technology, Shanghai 200093, China}
\affiliation{Shanghai Qi Zhi Institute, Shanghai 200232, China}
 \affiliation{Laboratory for Computational Physical Sciences (MOE),
 State Key Laboratory of Surface Physics, and Department of Physics,
 Fudan University, Shanghai 200433, China}

\author{Wenjing Xu}
\affiliation{College of Science, University of Shanghai for Science and Technology, Shanghai 200093, China}

\author{Di Lu}
 \affiliation{Laboratory for Computational Physical Sciences (MOE),
 State Key Laboratory of Surface Physics, and Department of Physics,
 Fudan University, Shanghai 200433, China}

\author{Yuxuan Zhou}
 \affiliation{Laboratory for Computational Physical Sciences (MOE),
 State Key Laboratory of Surface Physics, and Department of Physics,
 Fudan University, Shanghai 200433, China}

\author{Lu Liu}
 \affiliation{Laboratory for Computational Physical Sciences (MOE),
 State Key Laboratory of Surface Physics, and Department of Physics,
 Fudan University, Shanghai 200433, China}

 \author{Yaozhenghang Ma}
 \affiliation{Laboratory for Computational Physical Sciences (MOE),
 State Key Laboratory of Surface Physics, and Department of Physics,
 Fudan University, Shanghai 200433, China}

\author{Guangyu Wang}
 \affiliation{Laboratory for Computational Physical Sciences (MOE),
 State Key Laboratory of Surface Physics, and Department of Physics,
 Fudan University, Shanghai 200433, China}

\author{Hua Wu}
\email{Corresponding author. wuh@fudan.edu.cn}
\affiliation{Laboratory for Computational Physical Sciences (MOE),
 State Key Laboratory of Surface Physics, and Department of Physics,
 Fudan University, Shanghai 200433, China}
\affiliation{Shanghai Qi Zhi Institute, Shanghai 200232, China}
\affiliation{Collaborative Innovation Center of Advanced Microstructures,
 Nanjing 210093, China}

\date{\today}

\begin{abstract}
Hybrid transition metal oxides continue to attract attention due to their multiple degrees of freedom ($e.g.$, lattice, charge, spin, and orbital) and versatile properties. Here we investigate the magnetic and electronic properties of the newly synthesized double perovskite Ba$_2$NiIrO$_6$, using crystal field theory, superexchange model analysis, density functional calculations, and parallel tempering Monte Carlo (PTMC) simulations.
Our results indicate that Ba$_2$NiIrO$_6$ has the Ni$^{2+}$ ($t_{2g}^{6}e_{g}^{2}$)-Ir$^{6+}$ ($t_{2g}^{3}$) charge states. The first nearest-neighboring (1NN) Ni$^{2+}$-Ir$^{6+}$ ions prefer a ferromagnetic (FM) coupling as expected from the Goodenough-Kanamori-Anderson rules, which contradicts the experimental antiferromagnetic (AF) order in Ba$_2$NiIrO$_6$.
We find that the strong 2NN AF couplings are frustrated in the fcc sublattices, and they play a major role in determining the observed AF ground state. We also prove that the $J_{\rm eff}$ = 3/2 and $J_{\rm eff}$ = 1/2 states induced by spin-orbit coupling, which would be manifested in low-dimensional (e.g., layered) iridates, are however not the case for cubic Ba$_2$NiIrO$_6$. Our PTMC simulations show that when the long-range (2NN and 3NN) AF interactions are included, an AF transition with $T_{\rm N}$ = 66 K would be obtained and it is well comparable with the experimental 51 K. Meanwhile, we propose a possible 2$\times$2$\times$2 noncollinear AF structure for Ba$_2$NiIrO$_6$.
\end{abstract}

\maketitle


\section*{I. Introduction}

Transition metal (TM) perovskite oxides with the general formula $AB$O$_3$ represent a unique class of solids, where $A$ is an alkali, alkaline earth, or rare earth atom and $B$ is a transition metal. They exhibit fascinating properties and practical applications due to the interplay of charge, spin, orbital, and lattice degrees of freedom, such as high temperature ferromagnetic (FM) insulator, multiferroicity, colossal magnetoresistance, and superconductors~\cite{tokura2000,meng2018,khomskii2009,Tokura2006,orenstein2000}. In recent years, there has been an increasing interest in hybrid 3$d$, 4$d$, or 5$d$ TM double perovskites $A$$_2$$BB$$'$O$_6$~\cite{Kobayashi1998,huang2006,Khomskii2001}, where $B$ and $B$$'$ have a rock salt ordering forming an interweaved fcc sublattice. In such systems, the interplay of localized 3$d$ electrons and relatively delocalized 4$d$ or 5$d$ electrons offers more flexibility to produce extraordinary properties. For example, Sr$_2$FeMoO$_6$ has a large magnetoresistance with high $T_{\rm C}$ of about 410 K~\cite{Kobayashi1998}, and Sr$_2$CrB$'$O$_6$ (B$'$ = W/Re/Os) family shows a rapidly increasing $T_{\rm C}$ (from Sr$_2$CrWO$_6$ with 458 K~\cite{Philipp2003} via Sr$_2$CrReO$_6$ with 635 K~\cite{Kato2003} to Sr$_2$CrOsO$_6$ with 725 K~\cite{Krockenberger2007}). Moreover, 4$d$ and 5$d$ TMs have a strong spin-orbit coupling (SOC) which can bring about novel physical phenomena~\cite{Khomskii2001}, $e.g.$ a spin-orbit Mott state in Sr$_2$IrO$_4$~\cite{kim2009} and a spin-orbit controlled $J_{\rm eff}$ = 3/2 electronic ground state for 5$d^3$ systems Ca$_{3}$LiOsO$_{6}$ and Ba$_2$YOsO$_{6}$~\cite{Taylor2017}.

 \begin{figure}[H]
 \centering
\includegraphics[width=5.5cm]{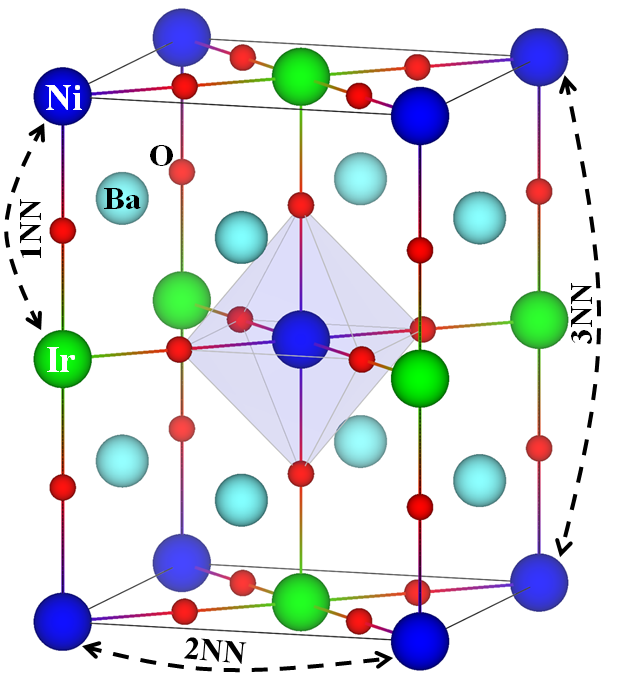}
\centering
\caption{The $\sqrt{2}a$$\times$$\sqrt{2}a$$\times$2$a$ structure for Ba$_2$NiIrO$_6$. 1NN, 2NN, and 3NN stand for the first, second, and third nearest-neighboring TMs, respectively.
}
\label{Structure}
\end{figure}

In the insulating perovskite oxides, their magnetic properties are very often understood by the superexchange coupling of the neighboring magnetic ions via intermediate oxygen, according to the Goodenough-Kanamori-Anderson (GKA) rules~\cite{Anderson1950,Goodenough1958,Kanamori1959}. For example, a linear TM-O-TM bond, with TM $d^8$-$d^3$ electronic configuration in an octahedral crystal field, is expected to be FM, and indeed La$_2$NiMnO$_6$ is a FM semiconductor with $T_{\rm C}$ $\sim$ 280 K and it has Ni$^{2+}$ 3$d^{8}$ and Mn$^{4+}$ 3$d^3$~\cite{Rogado2005}.
Very recently, the hybrid 3$d$-5$d$ double perovskite Ba$_2$NiIrO$_6$ is synthesized by high pressure and high temperature~\cite{Feng2021}, and it has a Ni-Ir ordered rock salt structure with the space group $Fm\bar{3}m$, see FIG. \ref{Structure}. The nominal Ni$^{2+}$-Ir$^{6+}$ (3$d^8$-5$d^3$) charge state is confirmed by an x-ray absorption near-edge spectroscopy. Somewhat surprisingly, Ba$_2$NiIrO$_6$ with the $d^8$-$d^3$ configuration displays an AF transition with N\'eel temperature ($T_{\rm N}$) of 51 K, in sharp contrast to the above expectation of FM state.
However, the fitting positive Weiss temperature of 80 K indicates a remarkable FM interaction, and the effective magnetic moment of 4.67 $\mu_{\rm B}$ per formula unit (fu) was measured~\cite{Feng2021}.

\begin{figure}[t]
  \centering
\includegraphics[width=8cm]{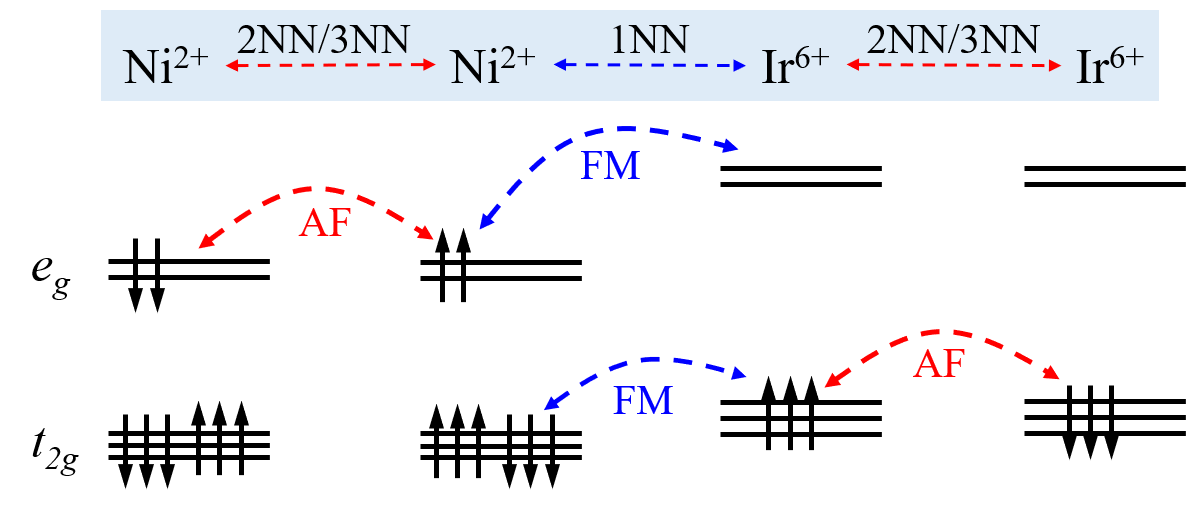}
\centering
 \caption{Schematic crystal field level diagrams of Ni$^{2+}$ and Ir$^{6+}$. Virtual electron hoppings from Ni$^{2+}$ to 1NN Ir$^{6+}$ would yield a FM coupling, while 2NN/3NN Ni$^{2+}$-Ni$^{2+}$ (and Ir$^{6+}$-Ir$^{6+}$) would be AF.
}
\label{states}
\end{figure}

The above controversy motivates us to study the electronic structure and magnetism of Ba$_2$NiIrO$_6$. In particular, we care about whether possible long-range magnetic interactions~\cite{ou2015,Kermarrec2015,Taylor2015,ou2018} of the delocalized $5d$ electrons and their strong SOC effects play a vital role in the intriguing AF ground state. As seen below, our results confirm the Ni$^{2+}$ ($t_{2g}^{6}$$e_{g}^{2}$)-Ir$^{6+}$ ($t_{2g}^{3}$) charge state. In a possible charge fluctuation process into a normal Ni$^{3+}$-Ir$^{5+}$ state rather than an abnormal Ni$^{+}$-Ir$^{7+}$, both superexchange channels from the virtual hopping of the Ni $e_g$ up-spin and $t_{2g}$ down-spin electrons would give rise to a FM coupling, see FIG. \ref{states}. The 1NN Ni$^{2+}$-Ir$^{6+}$ FM coupling is consistent with the GKA rules but contradicts the observed AF in Ba$_2$NiIrO$_6$. Therefore, the long-range magnetic couplings such as the 2NN and 3NN Ir-Ir couplings (and Ni-Ni ones) are worth a serious consideration, which are expected to be AF due to the half-filled Ir$^{6+}$ $t_{2g}^{3}$ and Ni$^{2+}$ $e_{g}^{2}$ shells, see FIG. \ref{states}. All these analyses are confirmed by our density functional theory (DFT) calculations. Then the 2NN AF couplings would be frustrated in the fcc sublattice, and they together with the 3NN AF play a major role in the experimental AF order, as seen in the following parallel tempering Monte Carlo (PTMC) simulations. Moreover, we find that the high coordination (12 Ir-Ir pairs) of the fcc Ir sublattice (and band formation of the delocalized Ir $5d$ electrons) and the Ir$^{6+}$ $t_{2g}^{3}$ exchange splitting both make the SOC ineffective. Thus, Ba$_2$NiIrO$_6$ can be treated as a spin-only system with a magnetic frustration, and it is predicted to have a possible 2$\times$2$\times$2 noncollinear AF structure.

%

\section*{II. Computational Details}
We perform density functional theory (DFT) calculations using the full-potential augmented plane waves plus local orbital code (Wien2k)~\cite{WIEN2K}.
The experimental cubic structure of Ba$_2$NiIrO$_6$ is used in our calculations, and the space group is $Fm\bar{3}m$ with $a$ = $b$ = $c$ = 4.02 $\angstrom$~\cite{Feng2021}.
To estimate the magnetic exchange parameters, a $\sqrt{2}a$$\times$$\sqrt{2}a$$\times$2$a$ supercell has been used in our calculations, see FIG. \ref{Structure}.
The muffin-tin sphere radii are chosen to be 2.8, 2.1, 2.1, and 1.5 Bohr for Ba, Ni, Ir, and O atoms, respectively.
The plane-wave cut-off energy of 16 Ry is set for the interstitial wave functions, and a $7\times7\times4$ $k$-mesh is used for integration over the Brillouin zone.
To describe the on-site electron correlation, the local spin density approximation plus Hubbard $U$ (LSDA+$U$) method\cite{Anisimov1993} is used, with the typical values of Hubbard $U$ = 6 eV (2 eV) and Hund exchange $J_{\rm H}$ = 1 eV (0.4 eV) for Ni 3$d$ (Ir 5$d$) electrons.
The spin-orbit coupling (SOC) is included for Ni and Ir atoms by the second-variational method with scalar relativistic wave functions.
The magnetic phase transition of Ba$_2$NiIrO$_6$ is probed using PTMC simulations~\cite{Hukushima1996} on a 16$\times$16$\times$16 spin matrix with periodic boundary conditions, and the number of replicas is set to 112.
Similar result is obtained with larger supercells. During the simulation step, each spin is rotated randomly in the three-dimensional space. The spin dynamical process is studied by the classical Metropolis methods~\cite{Metropolis}.
\section*{III. Results and Discussion}

 \begin{figure}[t]
\includegraphics[width=8cm]{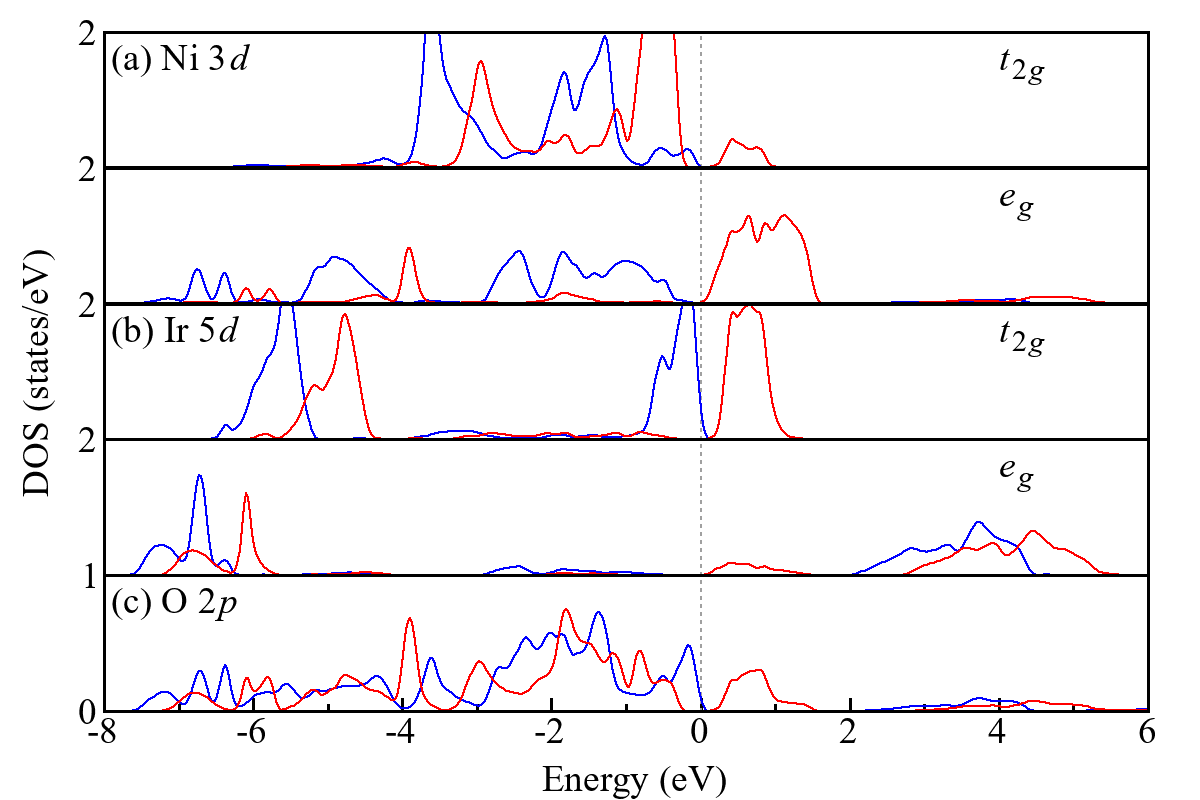}
\centering
 \caption{(a) The formal Ni$^{2+}$ 3$d^8$, (b) Ir$^{6+}$ 5$d^3$, and (c) O 2$p$ DOS for Ba$_2$NiIrO$_6$ by LSDA. The blue (red) curve stands for the up (down) spin. The Fermi level is set at zero energy.
}
\label{LSDA_DOS}
\end{figure}

We first perform the spin polarized LSDA calculations to study the charge state and the electronic structure of Ba$_2$NiIrO$_6$.
We plot in FIG. \ref{LSDA_DOS} the orbitally resolved density of states (DOS) for the FM state.
We find a large bonding-antibonding splitting in the Ir ion due to the significant Ir-O hybridization. As seen in FIG. \ref{LSDA_DOS}(b), the Ir $e_g$ ($t_{2g}$) electrons have a strong pd$\sigma$ ($pd$$\pi$) splitting up to 10 eV (6 eV) in this octahedral crystal-field.
The crystal-field splitting of Ir $e_g$-$t_{2g}$ states is about 3 eV. Except for the occupied bonding states (around --6 eV) assigned to the O 2$p$ bands, only the up-spin Ir $t_{2g}$ states are occupied and have an exchange splitting of about 1 eV, giving a formal Ir$^{6+}$ $t_{2g}^{3}$ configuration. In contrast, the Ni 3$d$ electrons have a smaller pd$\sigma$ ($pd$$\pi$) bonding-antibonding of about 4 eV (1 eV), and the $e_g$-$t_{2g}$ crystal-field splitting is about 1 eV, see FIG. \ref{LSDA_DOS}(a). Only the down-spin $e_g$ states are unoccupied, suggesting the formal Ni$^{2+}$ $t_{2g}^{6}$$e_{g}^{2}$ configuration. Therefore, Ba$_2$NiIrO$_6$ has the Ni$^{2+}$-Ir$^{6+}$  charge state.
Moreover, the calculated Ni (Ir) local spin moment of 1.70 (1.39) $\mu_{\rm B}$ refers to the nominal Ni$^{2+}$ $S$ = 1 (Ir$^{6+}$ $S$ = 3/2)  state. The large reduction of Ir spin moment is due to
the strong covalency with the oxygen ligands.
The total spin moment of 5 $\mu_{\rm B}$/fu in the FM state agrees well with the ideal Ni$^{2+}$ $S$ = 1 and Ir$^{6+}$ $S$ = 3/2 charge states. Taking into account a covalency reduction, this charge-spin state with the effective spin moment ($\sqrt{4\times1\times2+4\times3/2\times5/2}$ = 4.80 $\mu_{\rm B}$/fu) well explains the experimental effective moment of 4.67 $\mu_{\rm B}$/fu~\cite{Feng2021}.

\begin{figure}[t]
\includegraphics[width=8cm]{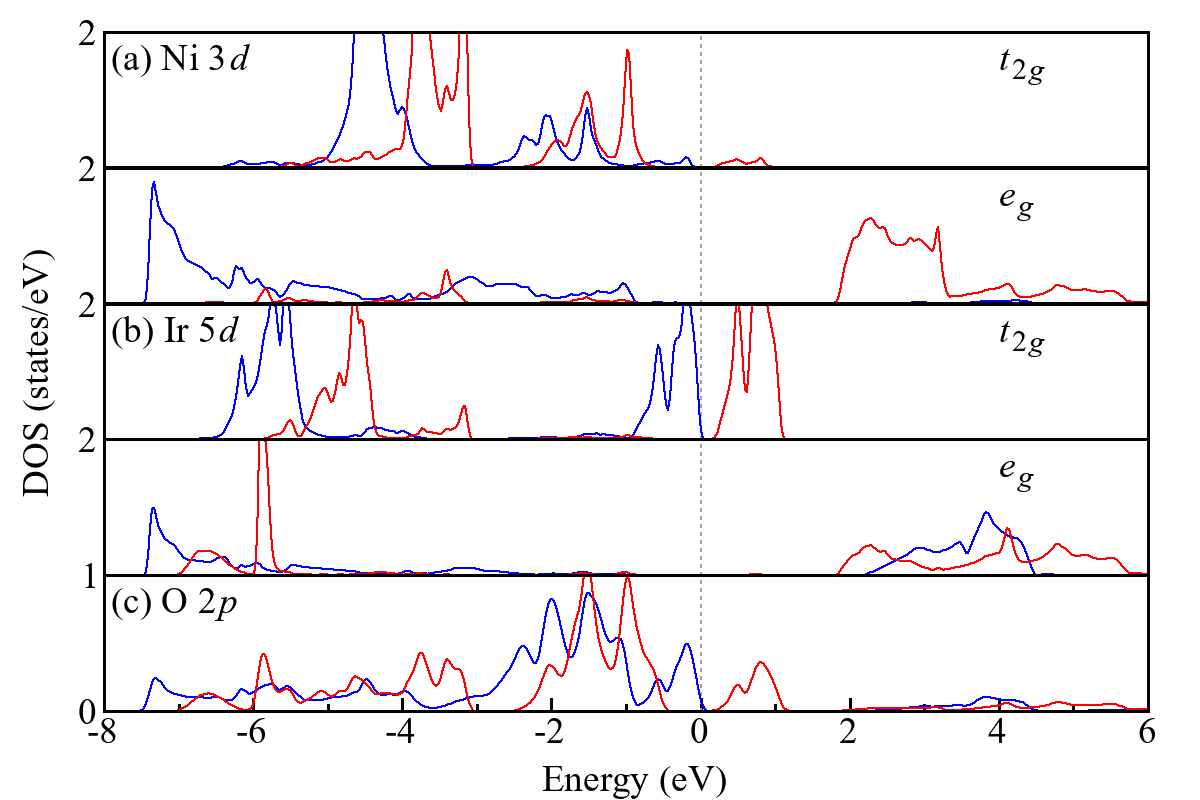}
\centering
 \caption{(a) The formal Ni$^{2+}$ 3$d^8$, (b) Ir$^{6+}$ 5$d^3$, and (c) O 2$p$ DOS for Ba$_2$NiIrO$_6$ by LSDA+$U$. The blue (red) curve stands for the up (down) spin. The Fermi level is set at zero energy.
}
\label{LSDA_U_DOS}
\end{figure}

To account for electronic correlations of the Ni and Ir ions, we carry out the LSDA+$U$ calculations. The enhanced electron localization gives rise to an increasing local spin moment of 1.77 $\mu_{\rm B}$ (1.42 $\mu_{\rm B}$) for Ni (Ir) ions, see TABLE \ref{tb1}. The total spin moment of 5 $\mu_{\rm B}$/fu is the same as the LSDA results, suggesting the formal Ni$^{2+}$ $S$ = 1 and Ir$^{6+}$ $S$ = 3/2 states again.
FIG. \ref{LSDA_U_DOS} shows the DOS results for the FM state calculated by LSDA+$U$. The energy level of the Ni unoccupied down-spin $e_g$ states becomes higher by comparison with FIG. \ref{LSDA_DOS}(a), due to the local Coulomb repulsion. In contrast, owing to the delocalization and weaker electron correlation, the Ir$^{6+}$ $5d$ states change insignificantly, by a comparison between FIG. \ref{LSDA_U_DOS}(b) and FIG. \ref{LSDA_DOS}(b).

\renewcommand\arraystretch{1.3}
\begin{table}[t]
\caption{Relative total energies $\Delta$\textit{E} (meV/fu), total spin moments ($\mu_{\rm B}$/fu), and local spin moments ($\mu_{\rm B}$) calculated by LSDA+\textit{U}. We assume artificial Ba$_2$ZnIrO$_6$ (La$_2$NiSiO$_6$) in Ba$_2$NiIrO$_6$ structure to estimate the Ir-Ir (Ni-Ni) exchange couplings.
The derived exchange parameters (meV) are listed in the last two rows.
}

\begin{tabular}{c@{\hskip3mm}c@{\hskip3mm}c@{\hskip3mm}c@{\hskip3mm}c@{\hskip3mm}}
\hline\hline
Systems        &  States & $\Delta$\textit{E} & \textit{M}$_{\rm TOT}$  & $\rm Ni^{2+}/Ir^{6+}$  \\ \hline
\multirow{2}{*}{Ba$_2$NiIrO$_6$}&FM       &     0       & 5.00         &    1.77  / 1.42   \\
               &G-AF     &     135     & --1.00    &    1.72  / --1.28      \\ \hline
\multirow{3}{*}{Ba$_2$ZnIrO$_6$}&  FM     &   0         &  3.00       &     / 1.39       \\
          &  Layered AF  &   --83      &   0.00      &     / 1.30         \\
          &  Bilayered AF&   --47      &   0.00     &     / 1.32        \\ \hline
\multirow{3}{*}{La$_2$NiSiO$_6$}&  FM     &    0        &   2.00      &   1.74  /          \\
          &  Layered AF  &    --18     &    0.00     &     1.73  /              \\
          &  Bilayered AF&     --9     &    0.00     &    1.73  /               \\ \hline\hline
 \textit{J}$_{\rm Ni-Ir}$&\textit{J}$'$$_{\rm Ir-Ir}$&\textit{J}$'$$_{\rm Ni-Ni}$&\textit{J}$''$$_{\rm Ir-Ir}$ &\textit{J}$''$$_{\rm Ni-Ni}$ \\
7.50 &--4.61 &--2.25 &--1.22  &0.00  \\ \hline\hline
 \end{tabular}
 \label{tb1}
\end{table}

As discussed above, both the Ni$^{2+}$ $S=1$ and Ir$^{6+}$ $S=3/2$ are magnetic in their interweaved fcc sublattices. Another significant aspect of Ba$_2$NiIrO$_6$ is the magnetic interactions. We plot in FIG. \ref{states} a schematic level diagram for Ni$^{2+}$ and Ir$^{6+}$ with 1NN/2NN/3NN magnetic interactions.
For the 1NN Ni$^{2+}$ and Ir$^{6+}$ ions, virtual hopping processes with an electronic excitation from Ni$^{2+}$ to Ir$^{6+}$ is possible, which gives rise to the excited state Ni$^{3+}$/Ir$^{5+}$. However, a reverse process into the highly unbalanced Ni$^{1+}$/Ir$^{7+}$ excited state should be strongly suppressed. Then, a FM Ni-Ir coupling is expected, associated with the charge excitation from Ni$^{2+}$-Ir$^{6+}$ to Ni$^{3+}$-Ir$^{5+}$, see FIG. \ref{states}. This is in line with the GKA rules. Ba$_2$NiIrO$_6$ would be a ferromagnet if only considering the 1NN Ni-Ir FM exchange, but this is strongly in conflict with the experimental AF order with $T_{\rm N}$ = 51 K~\cite{Feng2021}.
Thus, the long-range magnetic interactions of 2NN and even the 3NN should be taken into consideration. The far distanced Ir$^{6+}$-Ir$^{6+}$ coupling (and Ni$^{2+}$-Ni$^{2+}$) with the closed subshell favors AF, see FIG. \ref{states}.
Both Ir$^{6+}$ and Ni$^{2+}$ ions in double perovskite Ba$_2$NiIrO$_6$ form their respective fcc sublattices, thus giving rise to a magnetic frustration. Therefore, the magnetic properties of Ba$_2$NiIrO$_6$ are ultimately determined by the competition between the 1NN FM and 2NN/3NN AF couplings.

To estimate the exchange parameter of the 1NN Ni$^{2+}$-Ir$^{6+}$, we calculate two typical magnetic structures, $i.e$, FM and  G-AF (intralayer AF in $ab$ plane, interlayer AF along $c$ axis) within the LSDA+$U$ framework. Our results show that the FM is lower than the G-AF state by 135 meV/fu, see TABLE \ref{tb1}. Considering the exchange energy $J_{\rm Ni-Ir}$ for each 1NN Ni-Ir pair, the G-AF state differs from the FM state only by the 1NN Ni-Ir couplings, and thus the exchange energy difference per fu is expressed as
\begin{equation}
\begin{split}
E_{\rm G-AF} - E_{\rm FM} =& +12 \textit{J}_{\rm Ni-Ir}\textit{S}_{\rm Ni}\textit{S}_{\rm Ir}.
\end{split}
\end{equation}
Then 1NN Ni-Ir FM parameter $J_{\rm Ni-Ir}$ = 7.50 meV is derived. This FM parameter is consistent with the superexchange picture, see FIG. {\ref{states}}.

 \begin{figure}[t]
\includegraphics[width=8cm]{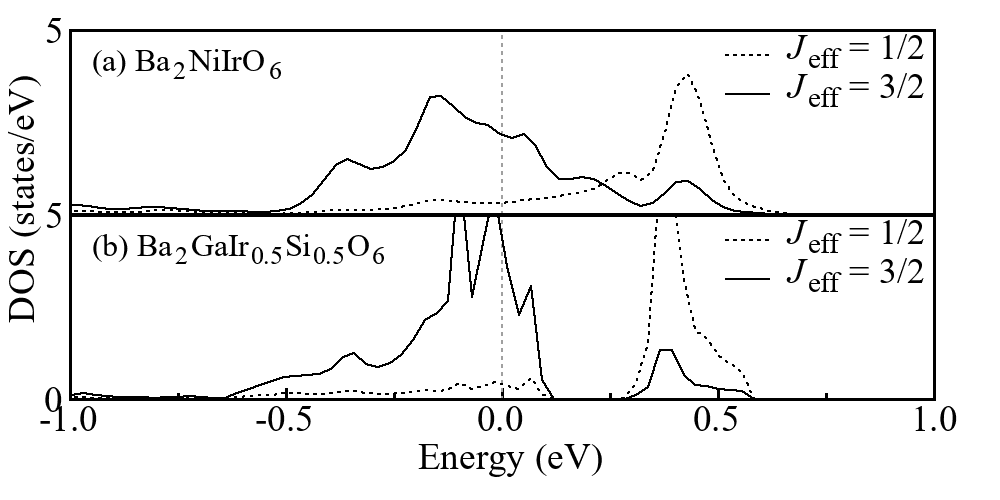}
\centering
 \caption{The Ir$^{6+}$ $t_{2g}$ DOS projected onto the SOC basis set by LDA+SOC. (a) In Ba$_2$NiIrO$_6$, the overall mixing of the  $J_{\rm eff}$ = 3/2 and $J_{\rm eff}$ = 1/2 states is due to the band formation of the delocalized Ir 5d electrons in the fcc Ir sublattice with 12 Ir-Ir coordination. (b) The SOC splitting between the $J_{\rm eff}$ = 3/2 and $J_{\rm eff}$ = 1/2 states is restored upon the reduction of the Ir-Ir coordination to four, which is modeled in the artificial system Ba$_{2}$GaIr$_{0.5}$Si$_{0.5}$O$_{6}$ with alternating GaIr and SiGa planes.
}
\label{LDA_SOC}
\end{figure}

To estimate the 2NN/3NN exchange parameters, two artificial systems Ba$_2$ZnIrO$_6$ and La$_2$NiSiO$_6$ are assumed in our calculations, see TABLE \ref{tb1}. Both Ba$_2$ZnIrO$_6$ and La$_2$NiSiO$_6$ are in the Ba$_2$NiIrO$_6$ structure, but only one fcc magnetic sublattice, either Ir or Ni, is present. This approach avoids the complicate magnetic structures in the otherwise bigger supercells of Ba$_2$NiIrO$_6$. For Ba$_2$ZnIrO$_6$, the layered AF and bilayered AF states are more stable than FM state by 83 meV/fu and 47 meV/fu, respectively, see TABLE \ref{tb1}. Note that both the layered AF and bilayered AF states are FM in $ab$ planes, but AF alternate (-up-down- or -up-up-down-down-) along the $c$ axis. Thus a $\sqrt{2}a$$\times$$\sqrt{2}a$$\times$4$a$ supercell is used to simulate the bilayered AF state, see FIG. \ref{Structure} for comparison. Considering the exchange energy $J'_{\rm Ir-Ir}$ ($J''_{\rm Ir-Ir}$) for each 2NN (3NN) Ir-Ir pair, the three different magnetic states have their respective exchange energy per fu as follows:
\begin{equation}
\begin{split}
E_{\rm FM}=  -6 \textit{J}'_{\rm Ir-Ir}\textit{S}_{\rm Ir}\textit{S}_{\rm Ir}
-3 \textit{J}''_{\rm Ir-Ir}\textit{S}_{\rm Ir}\textit{S}_{\rm Ir}\\
E_{\rm Layered-AF}=  +2 \textit{J}'_{\rm Ir-Ir}\textit{S}_{\rm Ir}\textit{S}_{\rm Ir}
-3 \textit{J}''_{\rm Ir-Ir}\textit{S}_{\rm Ir}\textit{S}_{\rm Ir}\\
E_{\rm Bilayered-AF}=  -2 \textit{J}'_{\rm Ir-Ir}\textit{S}_{\rm Ir}\textit{S}_{\rm Ir}
- \textit{J}''_{\rm Ir-Ir}\textit{S}_{\rm Ir}\textit{S}_{\rm Ir}.
\end{split}
\end{equation}
Then, the two AF parameters, $J'_{\rm Ir-Ir}$ = --4.61 meV and $J''_{\rm Ir-Ir}$ = --1.22 meV are derived, see TABLE \ref{tb1}. Similarly, the energy differences of the FM, layered AF, and bilayered AF states for La$_2$NiSiO$_6$ give AF $J'_{\rm Ni-Ni}$ = --2.25 meV and $J''_{\rm Ni-Ni}$ = 0.00 meV.
These results verify the superexchange pictures, see FIG. \ref{states}. Indeed, the 2NN Ir-Ir ions have a strong AF coupling, and the high coordination number of 12 in their fcc sublattice is twice that (6 coordination) of the 1NN FM Ni-Ir pairs. In addition, the relatively weak AF couplings exist in the 2NN Ni-Ni and 3NN Ir-Ir, and the 3NN Ni-Ni exchange is negligible. The stronger Ir-Ir couplings than the Ni-Ni ones are associated with the delocalization behavior of Ir $5d$ electrons versus the more localized Ni $3d$ electrons. Thus, Ba$_2$NiIrO$_6$ is a magnetically frustrated compound.

 \begin{figure}[t]
\includegraphics[width=8cm]{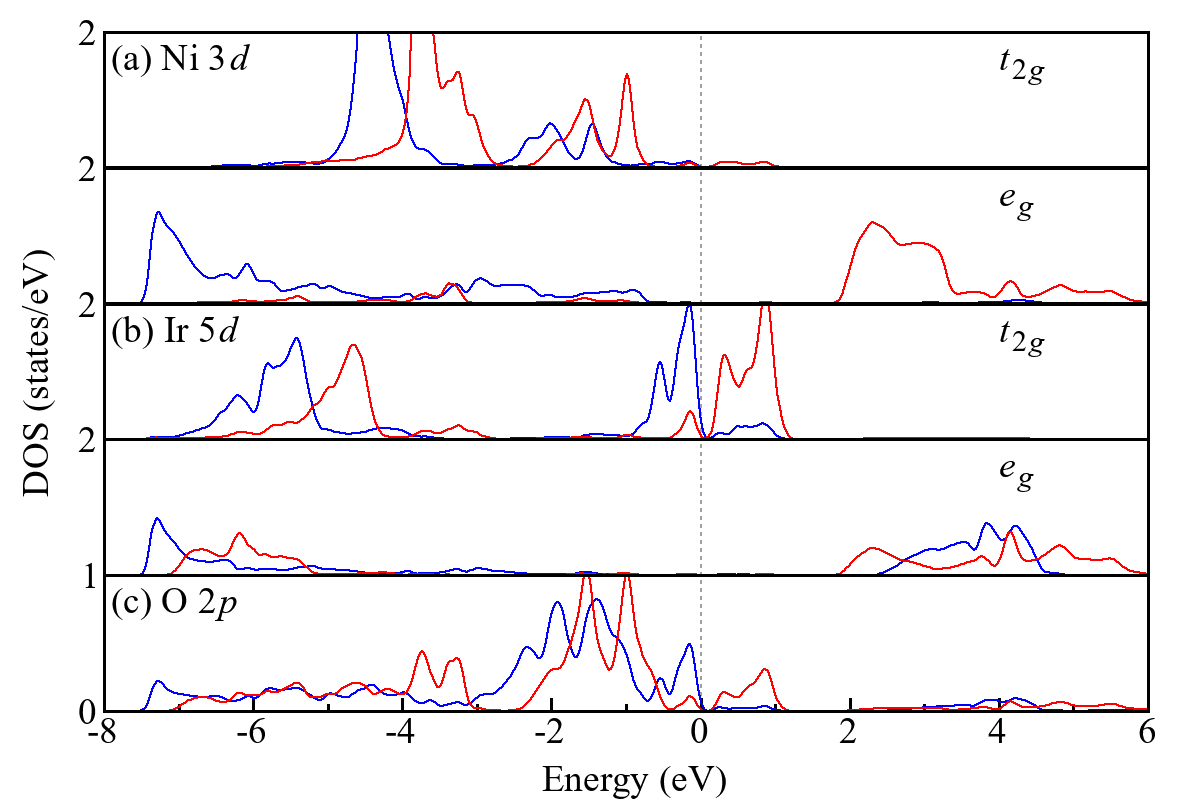}
 \caption{(a) Ni$^{2+}$ 3$d^8$, (b) Ir$^{6+}$ 5$d^3$, and (c) O 2$p$ DOS for Ba$_2$NiIrO$_6$ by LSDA+$U$+SOC. The blue (red) curve stands for the up (down) spin. The Fermi level is set at zero energy. Here the SOC turns out to be ineffective by a comparison with Fig. 4.
}
\label{LSDA_U_SOC_DOS}
\end{figure}

Spin-orbit coupling (SOC) in a heavy $5d$ transition metal is of a great concern, and particularly, iridates are under extensive studies~\cite{kim2009,Taylor2017,Guo2021,Feng2021,ou2015}. In the cubic crystal field, delocalized Ir 5$d$ orbitals see a large $e_{g}$-$t_{2g}$ splitting, and only the lower $t_{2g}$ orbitals are occupied. Taking the SOC effect into account, the $t_{2g}$ triplet (with 2-fold spin degeneracy) may split into the lower $J_{\rm eff}$ = 3/2 quartet and the higher $J_{\rm eff}$ = 1/2 doublet.
This SOC basis set is now used to project the Ir$^{6+}$ $t_{2g}$ orbitals of Ba$_2$NiIrO$_6$ in the LDA+SOC calculations. The clearly lower $J_{\rm eff}$ = 3/2 quartet and the higher $J_{\rm eff}$ = 1/2 doublet are observed, see FIG. \ref{LDA_SOC}(a). However, we find a strong mixing between them, suggesting that the $J_{\rm eff}$ = 3/2 and $J_{\rm eff}$ = 1/2 states are not good eigen orbitals for Ba$_2$NiIrO$_6$. This is due to the broad band formation of the delocalized Ir 5$d$ electrons within the fcc sublattice with the high coordination of 12.
In contrast, if the Ir-Ir coordination number is reduced, \textit{e.g.} as in the layered Sr$_{2}$IrO$_{4}$ under the planar 4 coordination, the SOC effect would be manifested.
To verify this, an artificial system Ba$_{2}$GaIr$_{0.5}$Si$_{0.5}$O$_{6}$ (also in Ba$_2$NiIrO$_6$ structure) with alternating GaIr and SiGa planes is calculated by LDA+SOC. Then, the planar Ir$^{6+}$-Ir$^{6+}$ 4 coordination is achieved by those ionic substitutions. As a result, an energy splitting of about 0.4 eV is obtained between the $J_{\rm eff}$ = 3/2 and $J_{\rm eff}$ = 1/2 states as seen in FIG. \ref{LDA_SOC}(b), and the Ir$^{6+}$ $t_{2g}^{3}$ electrons could occupy only the lower $J_{\rm eff}$ = 3/2 quartet.
Although the $J_{\rm eff}$ = 3/2 and $J_{\rm eff}$ = 1/2 states induced by the strong SOC appear in the low-dimensional iridates, they are not good eigen orbitals in the highly coordinated Ba$_2$NiIrO$_6$ with a broad band formation in the 12 coordinated fcc Ir sublattice.

We also perform the LSDA+$U$+SOC calculations by initializing the corresponding occupation number matrix with $t_{2g}^{3}$ $J$$_{\rm eff}$ = 3/2 states. After a full electronic relaxation, the results converge to the aforementioned Ni$^{2+}$ $S=1$ and Ir$^{6+}$ $S=3/2$ states. The obtained DOS results are almost the same as the LSDA+$U$ ones, see FIG. \ref{LSDA_U_SOC_DOS} and FIG. \ref{LSDA_U_DOS} for comparison. This indicates that indeed the SOC effect is negligible in Ba$_2$NiIrO$_6$. This is due to the band formation and to the larger Ir$^{6+}$ $t_{2g}^3$ exchange splitting of about 1 eV [Fig.~\ref{LSDA_DOS}(b)] than the Ir $5d$ SOC splitting of about 0.4 eV [Fig.~\ref{LDA_SOC}(b)]. Then, Ba$_2$NiIrO$_6$ can be treated as a spin-only system.

 \begin{figure}[t]
\includegraphics[width=8cm]{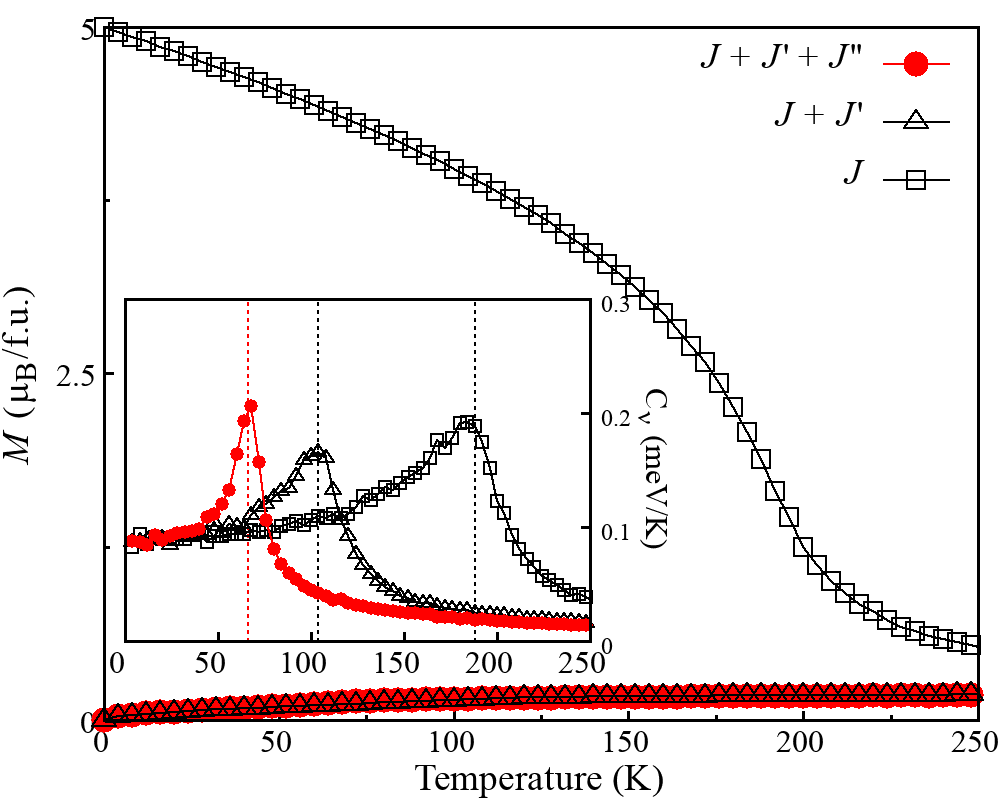}
\centering
 \caption{PTMC simulations of the magnetization of Ba$_2$NiIrO$_6$ as a function of temperature, by considering the different 1NN-FM/2NN-AF/3NN-AF couplings. The inset shows the magnetic specific heat.
}
\label{MC}
\end{figure}

\begin{figure}[t]
\includegraphics[width=8.5cm]{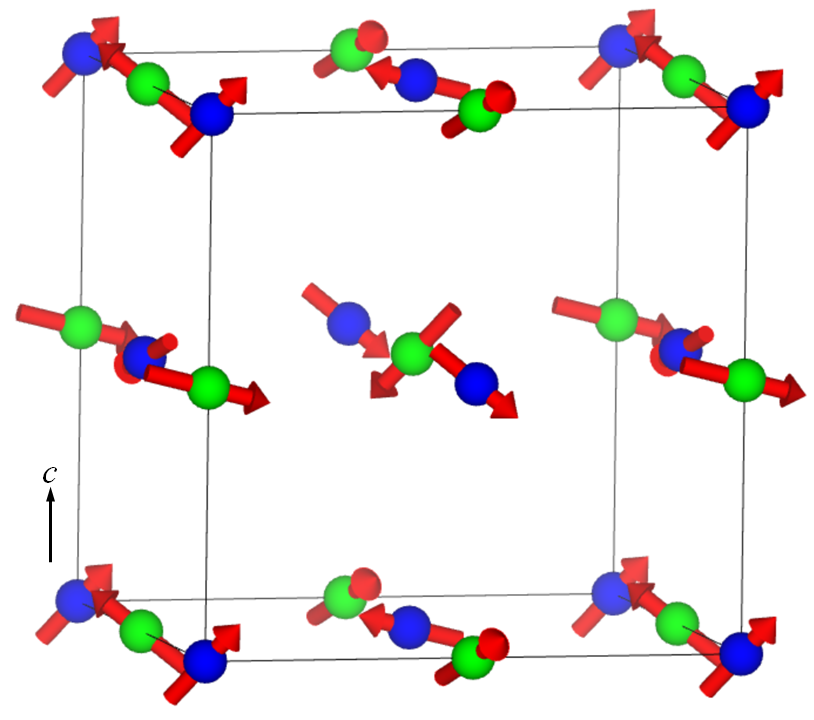}
\centering
 \caption{The possible $2\times2\times2$ noncollinear magnetic structure of Ba$_2$NiIrO$_6$ by PTMC simulations. The blue (green) balls represent for the Ni (Ir) atoms, and the Ba and O atoms are hidden for brevity. The red arrows stand for the spins.
}
\label{spin}
\end{figure}

We now perform PTMC simulations to see the impact of the long-range magnetic interactions on the magnetic structure of Ba$_2$NiIrO$_6$. Here, we assume the spin Hamiltonian
\begin{align}
\begin{split}
H = & -\frac{J}{2} \sum_{i,j}{\overrightarrow{S^{\rm Ni}_{i}} \cdot \overrightarrow{S^{\rm Ir}_{j}}}\\
&-\sum_{\alpha= \rm Ni,Ir}(\frac{J'}{2} \sum_{i,j}{\overrightarrow{S^{\rm \alpha}_{i}} \cdot \overrightarrow{S^{\rm \alpha}_{j}}}+\frac{J''}{2} \sum_{i,j}{\overrightarrow{S^{\rm \alpha}_{i}} \cdot \overrightarrow{S^{\rm \alpha}_{j}}})
\end{split}
\end{align}
in which the first term describes the 1NN FM Ni-Ir couplings, and the second term stands for the 2NN AF Ir-Ir (and Ni-Ni) couplings, and third term the 3NN AF Ir-Ir couplings (the 3NN Ni-Ni coupling being negligible). Using those exchange parameters listed in TABLE \ref{tb1}, our PTMC simulations show that the $T_{\rm C}$ would be 188 K with the total spin moment of 5 $\mu_{\rm B}$/fu when the 1NN Ni$^{2+}$-Ir$^{6+}$ FM exchange is included only, see Fig. \ref{MC}.
However, an AF behavior with the zero spin moment is obtained when the 2NN AF interactions are added, and the $T_{\rm N}$ would be 104 K. Moreover, when the 3NN AF Ir-Ir couplings are also included, the $T_{\rm N}$ would be further reduced to 66 K, and it is well comparable to the experimental $T_{\rm N}$ = 51 K in Ba$_2$NiIrO$_6$~\cite{Feng2021}.
Therefore, the strongest 1NN (local) FM Ni-Ir coupling accounts for the observed positive Weiss temperature, but the long-range AF interactions (particularly the 2NN and 3NN Ir-Ir couplings) bring about the magnetic frustration and eventually determine the low temperature AF order of Ba$_2$NiIrO$_6$~\cite{Feng2021}. Furthermore, we use PTMC simulations to probe the possible magnetic structure of the frustrated Ba$_2$NiIrO$_6$. Owing to the magnetic frustration, this system is hard to relax in MC simulations, and here we use the PTMC which is capable of driving the system to escape from the local minima at low temperatures. Then, a noncollinear 2$\times$2$\times$2 magnetic structure is obtained, see FIG. \ref{spin}, and it could be worth an experimental study.

Finally, we note that in many double perovskites containing the $5d$ transition metals, the long-range magnetic couplings are quite common and there could be complex magnetic structures due to the competing magnetic interactions. For example, Sr$_2$FeOsO$_6$ has two competing spin structures that differ in the spin sequence of ferrimagnetic Fe-Os layers~\cite{Paul2013}. Sr$_2$CoOsO$_6$ has the considerably stronger long-range superexchange interactions than the short Co-Os ones, and it has independent ordering of two interpenetrating magnetic sublattices~\cite{Morrow2013}, and the Co and Os sublattices exhibit different ground states and spin dynamics~\cite{Yan2014}. In contrast, Ca$_2$FeOsO$_6$ has a high-temperature ferrimagnetic order due to the strong Fe-Os superexchange which suppresses the magnetic frustration in the Fe and Os sublattices~\cite{Feng2014,Wang2014a}. All these findings suggest that both the short and long range magnetic couplings should be simultaneously invoked to account for the rich and complex magnetic structures of the hybrid transition metal oxides containing the $5d$ TMs.

\section*{IV. Summary}

In summary, using density functional calculations, crystal field level analyses, and PTMC simulations, we study the newly synthesized AF Ba$_2$NiIrO$_6$, which could be expected to have the Ni$^{2+}$ $3d^8$/Ir$^{6+}$ $5d^3$ FM interaction from the GKA rules. Our results show that the interplay between the long-range AF couplings and the 1NN FM one eventually leads to a frustrated magnetic structure. Our PTMC simulations reproduce the experimental AF transition temperature, and propose a possible 2$\times$2$\times$2 noncollinear AF structure. Moreover, we prove that here the SOC of Ir$^{6+}$ $5d^3$ shell is ineffective, due to the high coordination (12 Ir-Ir pairs) of the fcc Ir sublattice (and the broad band formation of the delocalized Ir $5d$ electrons) and the large exchange splitting of the half-filled Ir$^{6+}$ $t_{2g}^3$ shell. We note that Ba$_2$NiIrO$_6$ is a spin-only system, and that one could take care of both the short and long range magnetic interactions in the hybrid transition metal oxides containing the $5d$ TMs.

\section*{Acknowledgements}
This work was supported by National Natural Science Foundation of China (Grants No. 12104307 and No. 12174062).


\bibliographystyle{apsrev4-1}
\bibliography{BaNiIrO}

\end{document}